\documentclass[12pt]{article}
\usepackage{subfloat}
\pdfoutput = 1
\usepackage{graphics}
\usepackage{comment}
\usepackage{graphicx} %Include figure filesusepackage{graphicx} %Include figure files
\begin{document}
\date{Today}
%%%%%%%%%%%%%%%%%%%%
\title{{\bf{\Large  Analytical study of holographic superconductor with backreaction in $4d$ Gauss-Bonnet gravity}}}
%%%%%%%%%%%%%%%%%%%%

\author{
{\bf {\normalsize Debabrata Ghorai}$^{a}$
\thanks{debanuphy123@gmail.com, debabrataghorai@bose.res.in}},\,
{\bf {\normalsize Sunandan Gangopadhyay}$^{a}
$\thanks{sunandan.gangopadhyay@gmail.com, sunandan@iucaa.ernet.in}}\\
$^{a}$ {\normalsize  S.N. Bose National Centre for Basic Sciences,}\\{\normalsize JD Block, 
Sector III, Salt Lake, Kolkata 700106, India}\\[0.2cm]
%$^{b}$ {\normalsize Department of Physics, West Bengal State University, Barasat, India}\\
%$^{c}${\normalsize Visiting Associate in Inter University Centre for Astronomy \& Astrophysics,}\\
%{\normalsize Pune, India}\\[0.1cm]
}
\date{}

\maketitle

\begin{abstract}
{\noindent  In this paper we have analytically investigated holographic superconductors in four dimensional Einstein-Gauss-Bonnet gravity background. Recently the novel four dimensional Einstein-Gauss-Bonnet gravity has been formulated by rescaling the Gauss-Bonnet coupling constant $\alpha\rightarrow \frac{\alpha}{d-4}$ and taking the limit $d\rightarrow 4$, and this predicts several interesting new features. To know the effect of the curvature correction on the $(2+1)$-dimensional superconductors, we have consider the $3+1$-dimensional Einstein-Gauss-Bonnet gravity. Using the Sturm-Liouville eigenvalue approach, the critical temperature and the condensation operator values has been investigated in this framework away from the probe limit. We recover the numerical results from our analytical investigation.}
\end{abstract}
\vskip 1cm

%%%%%%%%%%%%%%%%%%%%%%%%%%%%%%%%% Introduction 
\section{Introduction}
The $AdS/CFT$ correspondence \cite{adscft1}-\cite{adscft4} relates strongly coupled systems living in the boundary to weakly coupled gravitational systems living in the bulk. It is well known that strongly coupled systems are difficult to explain using conventional tools, namely, perturbation theory approach. It is due to this fact that strongly coupled superconductors are analysed using its dual weakly coupled gravity model with the help of the $AdS/CFT$ dictionary. To explain the facts of high $T_c$ superconductors, one higher spatial dimensional gravity theory is considered, and thus the name `holographic superconductors'. The theoretical foundation of holographic superconductors \cite{hs6a} is based on the spontaneous symmetry breaking of the 
$U(1)$ symmetry \cite{hs1},\cite{hs2} of an Abelian Higgs model coupled to the gravity theory in $AdS$ spacetime. In the past decade there has been several studies on holographic superconductor models in presence of electrodynamics \cite{hs3}-\cite{hs21} and different gravity models \cite{hs22}-\cite{hs15}. The phase transition of superconductors is associated with the second order phase transition in gravity theory. To incorporate higher order curvature correction in the phase transition of gravity model, the Gauss-Bonnet gravity model is considered in higher dimensional spacetime ($d\geq 5$) since higher curvature correction does not contribute in any dynamical quantities in $d\leq 4$.

\noindent The Gauss-Bonnet (GB) gravity \cite{deser}-\cite{cai} has attracted a lot of attention among gravity theories with higher curvature corrections for $d\geq 5$. Mermin-Wagner theorem suggests that the phase transition may be affected due to higher curvature correction which leads to consider GB gravity in the context of phase transition in gravity theory. It is well known that the equation of motion for gravity in $4$-dimensional spacetime does not get affected due to the GB term in the Einstein-Gauss-Bonnet (EGB) gravity action. This is because the GB term yields a total derivative term when the variation of the EGB action in $4$-dimensional spacetime is carried out. Hence it does not contribute to the Einstein field equations and therefore does not contribute to any physical quantity in $4$-dimensional spacetime. For theoretical purpose, GB gravity for $d\geq 5$ dimensional spacetime has been studied for a long time. Recently, it has been found that higher curvature correction theorem may hold even in $4$-dimensional spacetime and predicts several new features which can be observed in the near future to test the theory. The novel four-dimensional EGB gravity with the singularity resolution for spherically symmetric solutions has been reported in \cite{4dgb1} by rescaling the GB coupling parameter $\alpha$ to $\frac{\alpha}{d-4}$ and taking the limit $d\rightarrow 4$ in $d$-dimensional EGB gravity action. The charged AdS black-hole solution is obtained for $4d$ EGB gravity \cite{4dgb2} and Einstein-Lovelock gravity \cite{4dgb3}, which is same as the solution of conformal anomaly inspired gravity \cite{4dgb4}. The various aspects of the $4d$ EGB gravity \cite{4dgba1}-\cite{4dgba4} has been investigated including gravitational lensing \cite{lencing1},\cite{lencing2}, the thermodynamics and cosmic censorship conjecture \cite{conjecture1},\cite{conjecture2}, the observable shadows\cite{shadow1},\cite{shadow2}, the quasinormal modes \cite{quasi} and stability\cite{stability1},\cite{stability2}. Some criticism on this theory were revealed in \cite{4dgbc1}-\cite{4dgbc4}. With the resolution of the criticism, a consistent realization of the $d\rightarrow 4$ limit of EGB gravity is proposed using the Arnowitt-Deser-Misner (ADM) decomposition in \cite{4dgbm}. Friedmann-Lemaître-Robertson-Walker (FLRW) and black hole solutions obtained
in \cite{4dgb1} are examined explicitly in \cite{4dgbch}.
%The grey-body factor for Dirac, electromagnetic and gravitational fields has been investigated and estimated the intensity of Hawking radiation and lifetime for asymptotically flat black holes in this consistent theory \cite{4dgbm1}. \\

\noindent The study of curvature corrections in $(3+1)$-dimensional holographic superconductors has been of some interest because of the Mermin-Wagner theorem which forbids continuous symmetry breaking in $(2+1)$-dimensions. The higher curvature correction does not show any effect on $3+1$-dimensional holographic superconductors. All studies of holographic dual model of superconductors in context of the effect of higher curvature correction are based on the framework of EGB gravity with $d\geq 5$. It is therefore interesting to investigate the effect of higher curvature corrections on $3+1$-dimensional holographic superconductors in the framework of the novel four-dimensional EGB gravity. Using this motivation, the properties of the $s$-wave and $p$-wave holographic superconductors has been studied numerically in the framework of 4d EGB gravity in the probe limit in \cite{jing}. The analytical investigation and the study away from the probe limit has not done yet for this holographic superconductor model.  

\noindent In this paper, using the Sturm-Liouville (SL) eigenvalue approach, we have analytically investigated the critical temperature and the condensation operator values away from the probe limit for $4d$ EGB gravity background. First, we have computed the $4d$ EGB metric solution without backreaction and with backreaction of the gauge field on the spacetime with our basic set up. To incorporate the backreaction on the metric, we have used the gauge field solution. We have then calculated the Hawking temperature from the metric solution, which is interpreted as the critical temperature of the boundary field theory. Near the critical temperature, we have recast the matter field equation in the SL eigenvalue form and estimated the eigenvalues with a trial function. We have then calculated the critical temperature in terms of the charge density. Using the perturbative expansion of the gauge field, we have computed the condensation operator values. We have shown that our analytical results match very well with the numerical findings.

%%%%%%%%%%%%%%%%%%%%%%%%%%%%%%%%%%%%%%%%%%%  

\noindent This paper is organized as follows. The basic set up for the $2+1$ holographic superconductors in 4d EGB gravity background has been discussed in section 2. In section 3, taking the backreaction of the matter fields on the spacetime metric, we have computed the metric solution for EGB gravity. In section 4, we have calculated the critical temperature in terms of charge density using SL eigenvalue method for $\psi_{-}=0,~\psi_{+}\neq 0$ case with and without backreaction. In section 5, the condensation operator value has been computed near the critical temperature. We have finally summarized our findings in section 6. 

%%%%%%%%%%%%%%%%%%%%%%%%%%%%%%%%%%%%%%%%%%%%%%%%%%%%%%%%%%%%%%%%%%%%
%%%%%%%%%%%%%%%%%%%%%%%%%%%%%  Section 2     

\section{Basic formalism }
Our basic starting point is to write down the action for the formation of scalar hair outside an electrically charged black hole in $d$-dimensional anti-de Sitter spacetime. This reads
\begin{eqnarray}
S=\int d^{d}x \frac{\sqrt{-g}}{2\kappa^{2}} \left( R -2\Lambda +\frac{\alpha}{2(d-4)}(R^{2} - 4 R^{\mu \nu} R_{\mu \nu} + R^{\mu \nu \lambda \rho } R_{\mu \nu \lambda \rho }) + 2 \kappa^2 \mathcal{L}_{matter} \right)
\label{ac1}
\end{eqnarray}
where $\Lambda=-(d-1)(d-2)/(2L^2)$ is the cosmological constant, $\kappa^2 = 8\pi G_{d} $ is the $d $-dimensional Newton's gravitational constant and $\alpha$ is the Gauss-Bonnet coupling parameter. The matter Lagrangian density reads
\begin{eqnarray}
\mathcal{L}_{matter}= -\frac{1}{4} F^{\mu \nu} F_{\mu \nu} - (D_{\mu}\psi)^{*} D^{\mu}\psi-m^2 \psi^{*}\psi
\label{ac2}
\end{eqnarray}
where $F_{\mu \nu}=\partial_{\mu}A_{\nu}-\partial_{\nu}A_{\mu}$ is the field strength tensor, 
$D_{\mu}\psi=\partial_{\mu}\psi-iqA_{\mu}\psi$ is the covariant derivative,  $A_{\mu}$ and $ \psi $ represent the gauge field and scalar field respectively.\\
The plane-symmetric black hole metric with backreaction can be written in the form
\begin{eqnarray}
ds^2=-f(r)e^{-\chi(r)}dt^2+\frac{1}{f(r)}dr^2+ r^2 h_{ij} dx^{i} dx^{j}
\label{m1}
\end{eqnarray}
where $ h_{ij} dx^{i} dx^{j}$ denotes the line element of a $(d-2) $-dimensional hypersurface with zero curvature. The Hawking temperature of this black hole is given by
\begin{eqnarray}
T_{H} = \frac{f^{\prime}(r_{+}) e^{-\chi(r_{+})/2}}{4\pi}
\label{gzx1}
\end{eqnarray} 
where $ r_{+} $ is the radius of the horizon of the black hole. This temperature is interpreted as the critical temperature of the boundary field theory. Now our task is to calculate this temperature in terms of the quantities of boundary field theory. To consider the effect of matter field on the spacetime, we first have to calculate the backreacted metric from the above action.  

\noindent Since we want to study the criticality of $s$-wave superconductor, the ansatz for the gauge field and the scalar field are \cite{hs6a}
\begin{eqnarray}
A = \phi(r)dt  ~~,~~\psi=\psi(r).
\label{vector}
\end{eqnarray}
The equations of motion for the metric and matter fields calculated with this ansatz read
\begin{eqnarray}
&&\left(1-\frac{(d-3) \alpha f(r)}{r^2}\right) f^{\prime}(r) + \frac{(d-3)f(r)}{r} - \frac{(d-1)r}{L^2}-\frac{(d-5)(d-3)}{2 r^3}\alpha f^2(r)  \nonumber\\
&& + \frac{2 \kappa^2 r}{d-2} \left[f(r)\psi^{\prime}(r)^2 + \frac{q^2 \phi^2(r) \psi^2(r) e^{\chi(r)}}{f(r)} + m^2 \psi^2(r) + \frac{\phi^{\prime}(r)^{2}e^{\chi(r)}}{2}\right]=0\nonumber \\
\label{e2}
\end{eqnarray}
\begin{eqnarray}
\left(1-\frac{(d-3) \alpha f(r)}{r^2}\right)\chi^{\prime}(r) + \frac{4 \kappa^2 r}{d-2}\left(\psi^{\prime}(r)^2 + \frac{q^2 \phi^2(r) \psi^2(r) e^{\chi(r)}}{f(r)^2}\right) = 0
\label{e02}
\end{eqnarray}
\begin{eqnarray}
\phi^{\prime \prime}(r) + \left(\frac{d-2}{r}+ \frac{\chi^{\prime} (r)}{2}\right) \phi^{\prime}(r) - \frac{2 q^2 \phi(r) \psi^{2}(r)}{f(r)} = 0
\label{e1}
\end{eqnarray}
\begin{eqnarray}
\psi^{\prime \prime}(r) + \left(\frac{d-2}{r}- \frac{\chi^{\prime} (r)}{2} + \frac{f^{\prime}(r)}{f(r)}\right)\psi^{\prime}(r) + \left(\frac{q^2 \phi^{2}(r) e^{\chi(r)}}{f(r)^2}- \frac{m^{2}}{f(r)}\right)\psi(r) = 0
\label{e01}
\end{eqnarray}
where prime denotes derivative with respect to $r$. 
The fact that $\kappa\neq0$ takes into account the backreaction of the spacetime. We now substitute $d=4$ for investigation of holographic superconductor in $4d$ Einstein-Gauss-Bonnet (EGB) gravity. Setting $L=1$, the equation of motion for metric and fields read from eq.(s)(\ref{e2}-\ref{e01}) in the limit $d\rightarrow 4$
\begin{eqnarray}
&&\left(1-\frac{\alpha f(r)}{r^2}\right) f^{\prime}(r)+\frac{f(r)}{r} - 3r+\frac{\alpha f^2(r)}{2 r^3}  \nonumber \\
&+& \kappa^2 r \left[f(r)\psi^{\prime}(r)^2 + \frac{q^2 \phi^2(r) \psi^2(r) e^{\chi(r)}}{f(r)} + m^2 \psi^2(r) + \frac{\phi^{\prime}(r)^{2}e^{\chi(r)}}{2}\right]=0 
\label{new3}
\end{eqnarray}
\begin{eqnarray}
\left(1-\frac{\alpha f(r)}{r^2}\right)\chi^{\prime}(r) + 2 \kappa^2 r\left(\psi^{\prime}(r)^2 + \frac{q^2 \phi^2(r) \psi^2(r) e^{\chi(r)}}{f(r)^2}\right) = 0
\label{new4}
\end{eqnarray}
\begin{eqnarray}
\phi^{\prime \prime}(r) + \left(\frac{2}{r}+ \frac{\chi^{\prime} (r)}{2}\right) \phi^{\prime}(r) - \frac{2 q^2 \phi(r) \psi^{2}(r)}{f(r)} = 0
\label{new5}
\end{eqnarray}
\begin{eqnarray}
\psi^{\prime \prime}(r) + \left(\frac{2}{r}- \frac{\chi^{\prime} (r)}{2} + \frac{f^{\prime}(r)}{f(r)}\right)\psi^{\prime}(r) + \left(\frac{q^2 \phi^{2}(r) e^{\chi(r)}}{f(r)^2}- \frac{m^{2}}{f(r)}\right)\psi(r) = 0 ~.
\label{new6}
\end{eqnarray}

\section{Metric for $4d$ Einstein-Gauss-Bonnet gravity}
\subsection{Without backreaction}
\noindent In $d=4$ spacetime, we now calculate the metric solution without backreaction $\kappa =0$. The metric equation (\ref{new3}) becomes 
\begin{eqnarray}
\left(1-\frac{\alpha f(r)}{r^2}\right) f^{\prime}(r) + \frac{f(r)}{r} - 3r+\frac{\alpha f^2(r)}{2 r^3} =0 ~.
\label{new1}
\end{eqnarray}
With the horizon condition ($f(r_{+}=0)$), the metric solution reads
\begin{eqnarray}
f(r) = \frac{r^2}{\alpha} \left[1- \sqrt{1-2\alpha\left(1-\frac{r^3_{+}}{r^3}\right)}\right]~. 
\label{new2}
\end{eqnarray} 
This is same as the metric in \cite{jing} with rescaling $\alpha$ to $2\alpha$.  
The asymptotic behaviour of the metric is 
\begin{eqnarray}
f(r) = \frac{r^2}{\alpha}\left[1- \sqrt{1-2\alpha}\right] \equiv \frac{r^2}{L^2_{eff}}
\label{new2a}
\end{eqnarray}
where $L_{eff}$ is the effective $AdS$ radius for $4d$ GB gravity which takes form 
\begin{eqnarray}
L_{eff} = \frac{\alpha}{1-\sqrt{1-2\alpha}}~.
\label{new2b}
\end{eqnarray}
From the above expression, we conclude that the upper limit of GB parameter is  $\alpha=\frac{1}{2}$ which is known as the Chern-Simons limit.

\subsection{With backreaction}
To incorporate the effect of matter on the spacetime, we have to consider backreacted metric. To obtain the backreacted metric due to the gauge field only, we need to solve the gauge field equation at $T=T_c$ in which matter field $\psi$ vanishes. We need the boundary conditions to know exact solution of gauge field. These boundary conditions are obtained from regularity condition of gauge field (i.e. $\phi(r_{+}=0)$) and the asymptotic behaviour of gauge field. In order to do this we need to fix the boundary conditions for $\phi(r)$ and $\psi(r)$ at the black hole horizon $r=r_+$ (where $f(r=r_+)=0$ with $e^{-\chi(r=r_+)}$ finite) and at the spatial infinity ($r\rightarrow\infty$) (where $f(\infty)=\infty$ with $e^{-\chi(r=r_+)}=1$). At boundary of spacetime, the field equation (\ref{new4}) reads
\begin{eqnarray}
\chi^{\prime}(r)= 0 ~.
\label{new7}
\end{eqnarray}
The conditions $e^{-\chi(r\rightarrow\infty)} \rightarrow 1 $, i.e. $\chi(r\rightarrow\infty)=0$ which in turn implies $\chi(r)=0$ from eq.(\ref{new7}). At boundary, the gauge field equation (\ref{new5}) reduces to 
\begin{eqnarray}
	\phi^{\prime\prime}(r) +\frac{2}{r} \phi^{\prime}(r)=0~.
\label{new8}
\end{eqnarray}
The solution of the gauge field reads
\begin{eqnarray}
	\phi(r)= c_1 -\frac{c_2}{r} ~.
\label{new9}
\end{eqnarray}
Using the AdS/CFT dictionary \cite{adscft2}-\cite{adscft3}, the constants $c_{1}$ and $c_2$ are identified with the chemical potential $\mu$ and the charge density $\rho$ respectively of the boundary field theory. Hence, the asymptotic behaviour of field can be written in terms of quantities of the boundary theory, which reads
\begin{eqnarray}
	\phi(r) = \mu - \frac{\rho}{r} ~.
\label{new10}
\end{eqnarray}
Similarly, the matter field eq.(\ref{new6}) takes the following form near the boundary of the spacetime ($f(r)\sim \frac{r^2}{L^2_{eff}}$ and $f^2(r)\sim \infty$)
\begin{eqnarray}
\psi^{\prime\prime}(r) + \frac{4}{r} \psi^{\prime}(r) - \frac{m^2 L^2_{eff}}{r^2} \psi(r)  =0 ~.
\label{new11}
\end{eqnarray}
The asymptotic behaviour of the matter field is (the solution of the above equation)
\begin{eqnarray}
	\psi(r) = \frac{\psi_{-}}{r^{\Delta_{-}}} +\frac{\psi_{+}}{r^{\Delta_{+}}}
\label{new12}
\end{eqnarray}
where $\Delta_{\pm} = \frac{1}{2}\left[3\pm \sqrt{9+4m^2 L^2_{eff}} \right]$ and $\psi_{\pm}$ represents the source or response term of the boundary theory. We observe that the Breitenlohner-Freedman bound \cite{bf1},\cite{bf2} $m^2_{BF} =-\frac{9}{4L^2_{eff}}$ depends on the GB parameter $\alpha$. We choose $\psi_{-}=0$ (or $\psi_{+}=0$) so that $\psi_{+}$ (or $\psi_{+}$) is dual to the expectation value of the condensation operator of the boundary theory in the absence of the source. \\

\noindent At the critical temperature $T=T_c$, $\psi(r)=0$, the field $\chi(r)=0$ and the gauge field eq.(\ref{new5}) reduces to
\begin{eqnarray}
\phi^{\prime\prime}(r) +\frac{2}{r} \phi^{\prime}(r)=0~.
\label{new13}
\end{eqnarray}
Using the regularization condition $(\phi(r_{+})=0)$ and the asymptotic behaviour of gauge field (\ref{new10}), the solution of the gauge field yields
\begin{eqnarray}
	\phi(r) = \lambda r_{+(c)} \left(1- \frac{r_{+(c)}}{r}\right) 
\label{new14}
\end{eqnarray} 
where $\lambda =\frac{\rho}{r^2_{+(c)}}$. The value of $\lambda$ will be determined by solving matter equation using the  Sturm-Liouvill eigenvalue method.  
Using eq.(\ref{new14}) and $\chi(r)=0$, the metric equation (\ref{new3}) at $T=T_c$ becomes
\begin{eqnarray}
\left(1-\frac{\alpha f(r)}{r^2}\right) f^{\prime}(r) + \frac{f(r)}{r} - 3r+\frac{\alpha f^2(r)}{2 r^3} + \frac{\kappa^2 \lambda^2 r^4_{+(c)}}{2r^3} =0 ~.
\label{new15}
\end{eqnarray}
Using the horizon condition $f(r_{+(c)})=0$, the backreacted metric takes form 
\begin{eqnarray}
f(r) = \frac{r^2}{\alpha} \left[1- \sqrt{1-2\alpha\left\{\left(1-\frac{r^3_{+(c)}}{r^3}\right)-\frac{\kappa^2\lambda^2}{2}\left(\frac{r^3_{+(c)}}{r^3}-\frac{r^4_{+(c)}}{r^4}\right)\right\}}\right] ~.
\label{new16}
\end{eqnarray}
For $\kappa=0$, we recover the metric (\ref{new2}) without backreaction.

%%%%%%%%%%%%%%%%%%%%%%%%%%%%%%%%%%%%%%%%%%%%%%%%%%%%%%%%%%%%%%%%%%%5

\section{The critical temperature} 
The Hawking temperature (\ref{gzx1}) for the backreacted metric (\ref{new16}) reads
\begin{eqnarray}
	T_{h} = \frac{r_{+}}{4\pi} \left[3-\frac{\kappa^2\lambda^2}{2}\right] ~~~~\Rightarrow~~~~ T_c = \frac{r_{+(c)}}{4\pi} \left[3-\frac{\kappa^2\lambda^2}{2}\right] =\frac{1}{4\pi} \left[3-\frac{\kappa^2\lambda^2}{2}\right]\frac{\sqrt{\rho}}{\sqrt{\lambda}}
\label{new17}
\end{eqnarray}
where we have substituted $r_{+(c)}$ from definition of $\lambda$.

\noindent Under the change of coordinates $z=\frac{r_{+(c)}}{r}$, the backreacted metric (\ref{new16}) therefore reads 
\begin{eqnarray}
f(z)= \frac{r^2_{+(c)}}{z^2} g_{0}(z)
\label{de7}
\end{eqnarray}
where 
\begin{eqnarray}
g_{0}(z)= \frac{1}{\alpha} \left[1- \sqrt{1-2\alpha\left\{(1-z^3)-\frac{\kappa^2\lambda^2}{2}(z^3-z^4)\right\}} \right] ~.
\label{metr33}
\end{eqnarray} 

\noindent In $z$-coordinate, the matter field (\ref{new6}) takes form near $T\rightarrow T_c$
\begin{eqnarray}
\psi''(z)+ \left(\frac{g'_{0}(z)}{g_{0}(z)}-\frac{2}{z}\right)\psi'(z)+ \left( \frac{\phi^{2} (z)}{g^{2}_{0} (z) r^{2}_{+(c)}} - \frac{m^2}{g_{0}(z) z^2}\right)\psi(z)=0
\label{e001}
\end{eqnarray}
where $\phi(z)$ now corresponds to the solution in eq.(\ref{new14}). In the above equation, we
shall also consider the fact that $\kappa^2 \lambda^2 =\kappa^2_i \lambda^2_{\kappa_{i-1}} $ which 
in turn implies that $g_{0}(z)$ reads as
\begin{eqnarray}
g_0(z) = \frac{1}{\alpha} \left[1- \sqrt{1-2\alpha\left\{(1-z^3)-\frac{\kappa^2_i\lambda^2_{\kappa_{i-1}}}{2}(z^3-z^4)\right\}} \right] ~.
\label{new20}
\end{eqnarray}
Near the boundary, we define \cite{siop}
\begin{eqnarray}
\psi(z)=\frac{\langle\mathcal{O}_{+}\rangle}{r^{\Delta_{+}}_{+(c)}} z^{\Delta_{+}} F(z)
\label{sol1}
\end{eqnarray}
where $F(0)=1$ and $\langle\mathcal{O}_{+}\rangle$ is the condensation operator.
Substituting this form of $\psi(z)$ in eq.(\ref{e001}), we obtain
\begin{eqnarray}
F''(z) &+& \left\{\frac{2\Delta_{+} -2}{z} +\frac{g'_{0}(z)}{g_{0}(z)} \right\}F'(z) +\left\{ \frac{\Delta_{+}(\Delta_{+} -1)}{z^2} +\left(\frac{g'_{0}(z)}{g_{0}(z)}-\frac{2}{z}\right)\frac{\Delta_{+}}{z}-\frac{m^2}{g_{0}(z) z^2} \right\}F(z) \nonumber \\
&+& \frac{\lambda^2}{g^{2}_{0}(z)} (1-z)^2 F(z)=0
\label{eq5b}
\end{eqnarray}
to be solved subject to the boundary condition $F' (0)=0$. 

\noindent It is now simple to see that the above equation can be written in the Sturm-Liouville form 
\begin{eqnarray}
\frac{d}{dz}\left\{p(z)F'(z)\right\}+q(z)F(z)+\lambda^2 r(z)F(z)=0
\label{sturm}
\end{eqnarray}
with 
\begin{eqnarray}
p(z)&=&z^{2\Delta_{+} -2}g_{0}(z)\nonumber\\
q(z)&=&z^{2\Delta_{+} -2}g_{0}(z)\left\{ \frac{\Delta_{+}(\Delta_{+} -1)}{z^2} +\left(\frac{g'_{0}(z)}{g_{0}(z)}-\frac{2}{z}\right)\frac{\Delta_{+}}{z}-\frac{m^2}{g_{0}(z) z^2} \right\} \nonumber\\
r(z)&=&\frac{z^{2\Delta_{+} -2}}{g_{0}(z)} (1-z)^2 ~. 
\label{i1}
\end{eqnarray}
The above identification enables us to write down an equation for the eigenvalue $\lambda^2$ which minimizes the expression 
\begin{eqnarray}
\lambda^2 &=& \frac{\int_0^1 dz\ \{p(z)[F'(z)]^2 - q(z)[F(z)]^2 \} }
{\int_0^1 dz \ r(z)[F(z)]^2}~.
\label{eq5abc}
\end{eqnarray}
We shall now use the trial function for the estimation of $\lambda^{2}$
\begin{eqnarray}
F= F_{\tilde\alpha} (z) \equiv 1 - \tilde\alpha z^2 ~.
\label{eq50}
\end{eqnarray}
Note that $F$ satisfies the conditions $F(0)=1$ and $F'(0)=0$.
The critical temperature from eq.(\ref{new17}) reads
\begin{eqnarray}
T_c = \frac{1}{4\pi} \left[3-\frac{\kappa^2_i\lambda^2_{\kappa_{i-1}}}{2}\right]\frac{\sqrt{\rho}}{\sqrt{\lambda}}
\label{new21}
\end{eqnarray}
where the Sturm-Liouville eigenvalue $\lambda$ is estimated from eq.(\ref{i1}-\ref{eq50}).\\

\noindent The mass of the scalar field can be selected from the value of $m^2L^2$ or $m^2L^2_{eff}$ which must satisfy the BF bound. The choice of $m^2L^2=-2$ or $m^2L^2_{eff}=-2$ yields $\Delta_{+}=2$.

\subsection{Without Backreaction}
Without backreaction ($\kappa =0$), the critical temperature takes form  
\begin{eqnarray}
T_c =\frac{3}{4\pi}\frac{\sqrt{\rho}}{\sqrt{\lambda}}
\label{new22}
\end{eqnarray}
with 
\begin{eqnarray}
g_0(z) = \frac{1}{\alpha} \left[1- \sqrt{1-2\alpha(1-z^3)} \right] ~.
\label{new23}
\end{eqnarray}
For $\alpha=0$, the above metric reduce to the metric for Einstein gravity $g_0(z) = 1-z^3$. First we consider $m^2L^2=-2$ case. Using eq.(\ref{i1}) and the trial function (\ref{eq50}), we find 
\begin{eqnarray}
\lambda_{\tilde\alpha}^2 = \frac{1-1.3333 \tilde{\alpha} + 0.80 \tilde{\alpha}^2}{0.043794 -0.0305579 \tilde{\alpha} + 0.007593 \tilde{\alpha}^2 } 
\label{est2}
\end{eqnarray}
which attains its minimum at $\tilde\alpha \approx 0.60159$. 
The critical temperature can now be computed from eq.(\ref{new22}) and reads 
\begin{eqnarray}
T_c =\frac{3}{4\pi\sqrt{\lambda|_{\tilde\alpha=.60159}}}\sqrt{\rho}\approx 0.117 \sqrt{\rho} 
\label{eqTc1}
\end{eqnarray}
which is in very good agreement with the numerical result
$T_c = 0.118 \sqrt{\rho}$ \cite{hs6a}.\\
We rerun the above analysis for $4d$ GB gravity  to estimate the minimum value of $\lambda^2$.  For the GB parameter $\alpha =0.05$, $\lambda^2$ attains its minimum value $\lambda^2 =19.3394$ at $\tilde{\alpha} \approx 0.593698$. From eq.(\ref{new22}), the critical temperature yields 
\begin{eqnarray}
T_c =\frac{3}{4\pi\sqrt{\lambda|_{\tilde\alpha=.593698}}}\sqrt{\rho}\approx 0.1138 \sqrt{\rho} 
\label{eqTc}
\end{eqnarray}
which is in very good agreement with the numerical result
$T_c = 0.114 \sqrt{\rho}$ \cite{jing}. For different values of the GB parameter, we have shown the critical temperature in Table \ref{tab1} which are in good agreement with numerical values in \cite{jing} ($\alpha$ is scaled by $2\alpha$ in \cite{jing}).\\
\begin{table}[h!]
\caption{For backreaction parameter $\kappa=0$ and $m^2L^2=-2$}
\centering
\begin{tabular}{|c| c| c| c| }
\hline
GB parameter & $\tilde\alpha $ & $\lambda^{2}$& $\frac{T_c}{\sqrt{\rho}}$ (analytical)  \\
\hline
$\alpha=$ -0.10 & 0.6145 & 13.928 & 0.1236\\
$\alpha=$ -0.05 & 0.6085 & 15.519 & 0.1202\\
\hline
$\alpha=$ 0.00 & 0.6016 & 17.309 & 0.1170  \\
\hline
$\alpha=$ 0.05 & 0.5937 & 19.339 & 0.1138 \\
$\alpha=$ 0.10 & 0.5845 & 21.666 & 0.1107 \\ 
$\alpha=$ 0.20 & 0.5602 & 27.542 & 0.1042 \\
$\alpha=$ 0.30 & 0.5219 & 36.050 & 0.0974 \\
$\alpha=$ 0.40 & 0.4474 & 50.109 & 0.0897 \\
\hline
$\alpha=$ 0.49 & 0.2083 & 79.445 & 0.0799 \\
$\alpha=$ 0.495 & 0.1622 & 83.178 & 0.0791 \\
$\alpha=$ 0.50 & 0.0703 & 89.083 & 0.0777 \\
\hline
\end{tabular}
\label{tab1}
\end{table} \\
\noindent We now calculate the critical temperature for $m^2L^2_{eff}=-2$ in which the mass of the scalar field depends on the GB parameter by eq.(\ref{new2b}). Rerunning the above procedure, we obtain the critical temperature for $m^2L^2_{eff}=-2$ which are shown in Table \ref{tab2} which are in good agreement with the numerical values in \cite{jing}.
\begin{table}[h!]
\caption{For backreaction parameter $\kappa=0$ and $m^2L^2_{eff}=-2$}
\centering
\begin{tabular}{|c| c| c| c| }
\hline
GB parameter & $\tilde\alpha $ & $\lambda^{2}$& $\frac{T_c}{\sqrt{\rho}}$ (analytical)  \\
\hline
$\alpha=$ -0.10 & 0.6088 & 16.019 & 0.1193\\
$\alpha=$ -0.05 & 0.6055 & 16.634 & 0.1182\\
\hline
$\alpha=$ 0.00 & 0.6016 & 17.309 & 0.1170  \\
\hline
$\alpha=$ 0.05 & 0.5971 & 18.058 & 0.1158 \\
$\alpha=$ 0.10 & 0.5918 & 18.893 & 0.1145 \\ 
$\alpha=$ 0.20 & 0.5775 & 20.905 & 0.1117 \\
$\alpha=$ 0.30 & 0.5547 & 23.582 & 0.1083 \\
$\alpha=$ 0.40 & 0.5097 & 27.371 & 0.1044 \\
\hline
$\alpha=$ 0.49 & 0.3652 & 30.789 & 0.1014 \\
$\alpha=$ 0.495 & 0.3380 & 30.035 & 0.1020 \\
$\alpha=$ 0.50 & 0.2944 & 24.027 & 0.1078 \\
\hline
\end{tabular}
\label{tab2}
\end{table} \\

\begin{figure}[t!]
\centering
\includegraphics[scale=0.4]{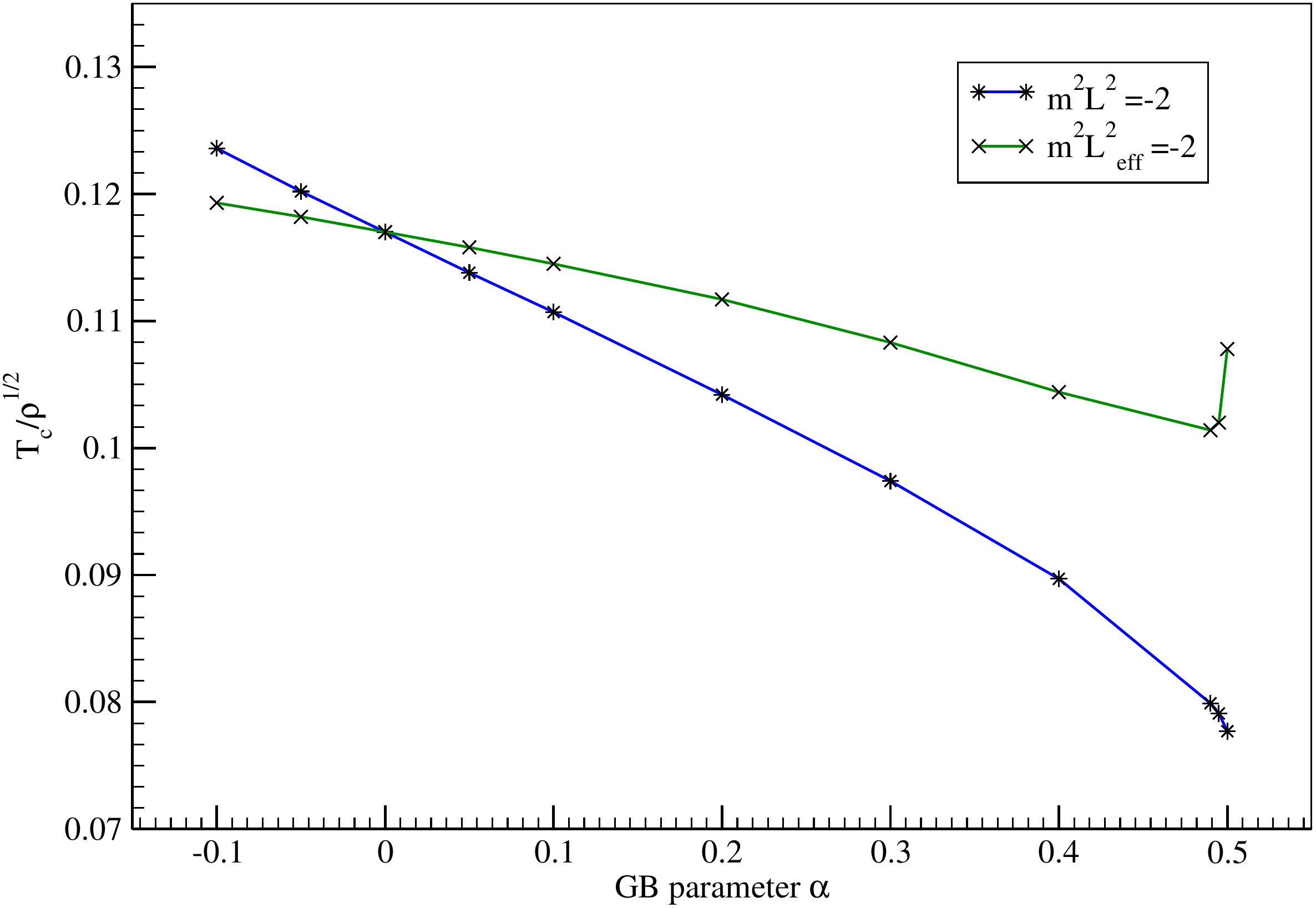} 
\caption{Plot of $\frac{T_c}{\sqrt{\rho}}$ vs. GB parameter $\alpha$ for $m^2L^2=-2$ and $m^2L^2_{eff}=-2$.}
\label{fig1}
\end{figure}

\noindent We have summarized our analytical results of the critical temperature with the GB parameter in Figure \ref{fig1}. Here we observe that the critical temperature decreases with increasing value $\alpha$ for both cases $m^2L^2=-2$ and $m^2L^2_{eff}=-2$. However, the critical temperature near the upper bound of GB parameter $\alpha$ (Chern-Simons limit), is suddenly increasing for $m^2L^2_{eff}=-2$ which is different from the case $m^2L^2=-2$.

%%%%%%%%%%%%%%%%%%%%%%%%%%%%%%%%%%%%%%%%%%%%%%%%%%%%%%%%%%%%%%%%%%%%%%%%%%%%%%%%%%%%%%%%%%%%%%%%%%%%%%%%%%%%%%%%%%
%%%%%%%%%%%%%%%%%%%%%%%%%%%%%%%%%%%%%%% %%%%%%%%%%%%%%%
\subsection{With Backreaction}
We now compute the critical temperature with backreacted metric (\ref{metr33}). 
For $\alpha=0$, the above metric reduce to the backreacted metric for Einstein gravity 
\begin{eqnarray}
g_0(z) = 1-z^3 -\frac{\kappa^2_i \lambda^2_{\kappa_{i-1}}}{2}(z^3- z^4) 
\end{eqnarray}
which is same as in \cite{hs15}. We concentrate here to compute the critical temperature for $4d$ GB gravity. We first analyse the case $m^2L^2=-2$. 
Using eq.(\ref{new20}) and eq.(\ref{i1}) with trial function (\ref{eq50}), we find  that $\lambda^2$ attains its minimum value $\lambda^2 =18.99$ at $\tilde{\alpha} \approx 0.58587$ for backreaction parameter $\kappa=0.1$ and $\alpha=0.05$. From eq.(\ref{new22}), the critical temperature yields 
\begin{eqnarray}
T_c =\frac{3}{4\pi\sqrt{\lambda|_{\tilde\alpha=.58587}}}\sqrt{\rho}\approx 0.1107 \sqrt{\rho} ~.
\label{eqTc}
\end{eqnarray}
For different value of backreaction parameter $\kappa$ and GB parameter $\alpha$, we have shown the critical temperature in Table \ref{tab3}.\\
\begin{table}[h!]
\caption{The values of $\frac{T_c}{\sqrt{\rho}}$ for different values of $\kappa$ and $\alpha$ in the case $m^2L^2=-2$}
\centering
\begin{tabular}{|c| c| c| c| c| c| }
\hline
$\frac{T_c}{\sqrt{\rho}}$ & $\alpha=0.05$ & $\alpha=0.10$ &  $\alpha=0.40$ & $\alpha=0.49$ & $\alpha=0.50$   \\
\hline
$\kappa=$ 0.1 & 0.1107 & 0.1072 & 0.0834 & 0.0715 & 0.0687 \\
\hline
$\kappa=$ 0.2 & 0.1013 & 0.0970 & 0.0653 & 0.0484 & 0.0446 \\ 
\hline
$\kappa=$ 0.3 & 0.0867 & 0.0813 & 0.0423 & 0.0299 & 0.0297 \\
\hline
$\kappa=$ 0.4 & 0.0691 & 0.0628 & 0.0268 & 0.0228 & 0.0193 \\
\hline
\end{tabular}
\label{tab3}
\end{table} \\
\noindent We now calculate the critical temperature for $m^2L^2_{eff}=-2$ in which the mass of the scalar field depends on the GB parameter by eq.(\ref{new2b}). Rerunning the above procedure, we obtain the critical temperature for $m^2L^2_{eff}=-2$ which are shown in Table \ref{tab4}.
\begin{table}[h!]
\caption{The values of $\frac{T_c}{\sqrt{\rho}}$ for different values of $\kappa$ and $\alpha$ in the case $m^2L^2_{eff}=-2$}
\centering
\begin{tabular}{|c| c| c| c| c| c| }
\hline
$\frac{T_c}{\sqrt{\rho}}$ & $\alpha=0.05$ & $\alpha=0.10$ &  $\alpha=0.40$ & $\alpha=0.49$ & $\alpha=0.50$   \\
\hline
$\kappa=$ 0.1 & 0.1128 & 0.1114 & 0.1006 & 0.0979 & 0.1056 \\
\hline
$\kappa=$ 0.2 & 0.1040 & 0.1023 & 0.0896 & 0.0880 & 0.0995 \\ 
\hline
$\kappa=$ 0.3 & 0.0902 & 0.0882 & 0.0743 & 0.0764 & 0.0930 \\
\hline
$\kappa=$ 0.4 & 0.0734 & 0.0713 & 0.0592 & 0.0676 & 0.0886 \\
\hline
\end{tabular}
\label{tab4}
\end{table} \\

\begin{figure}[t!]
\centering
\includegraphics[scale=0.3]{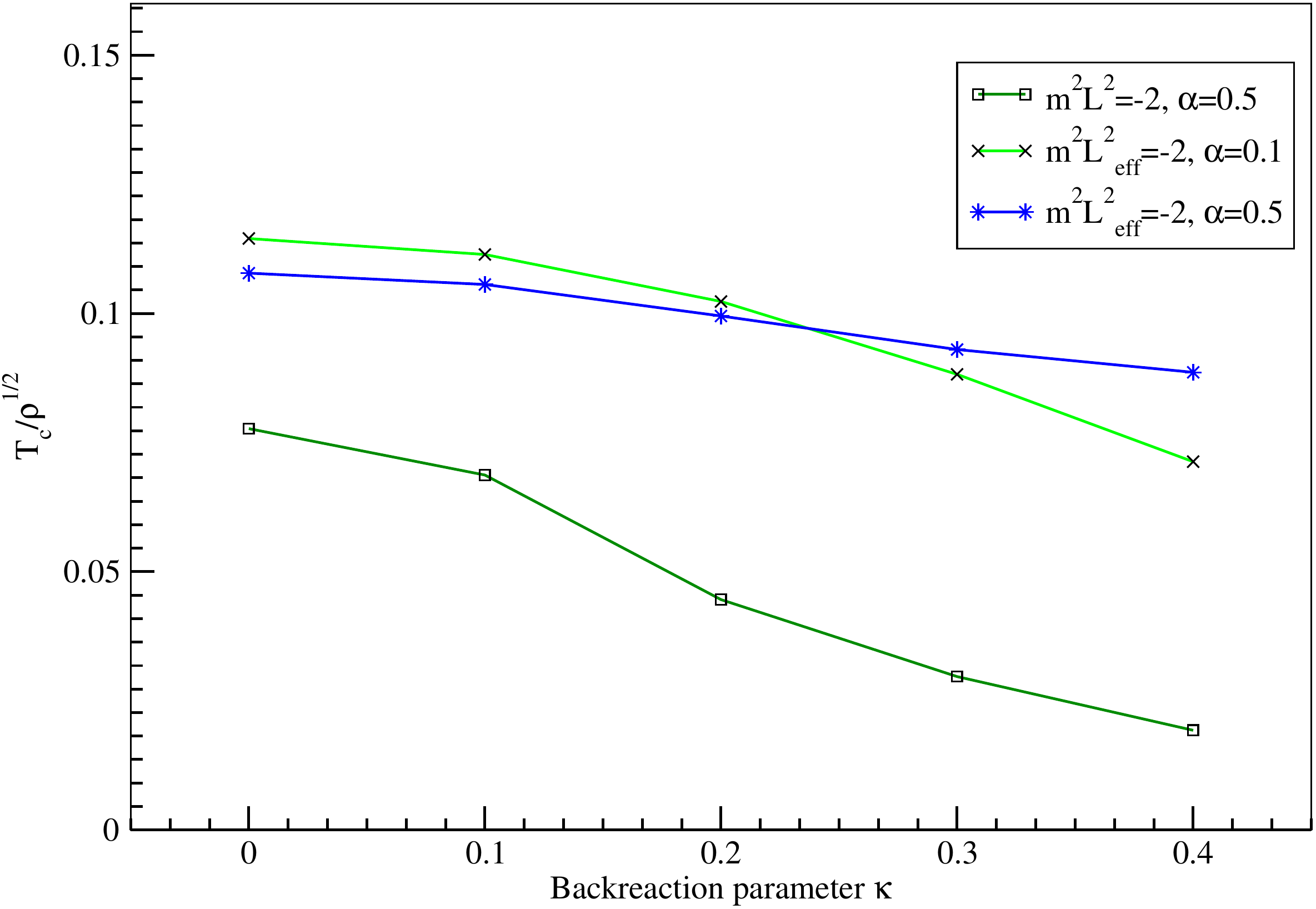} 
\includegraphics[scale=0.3]{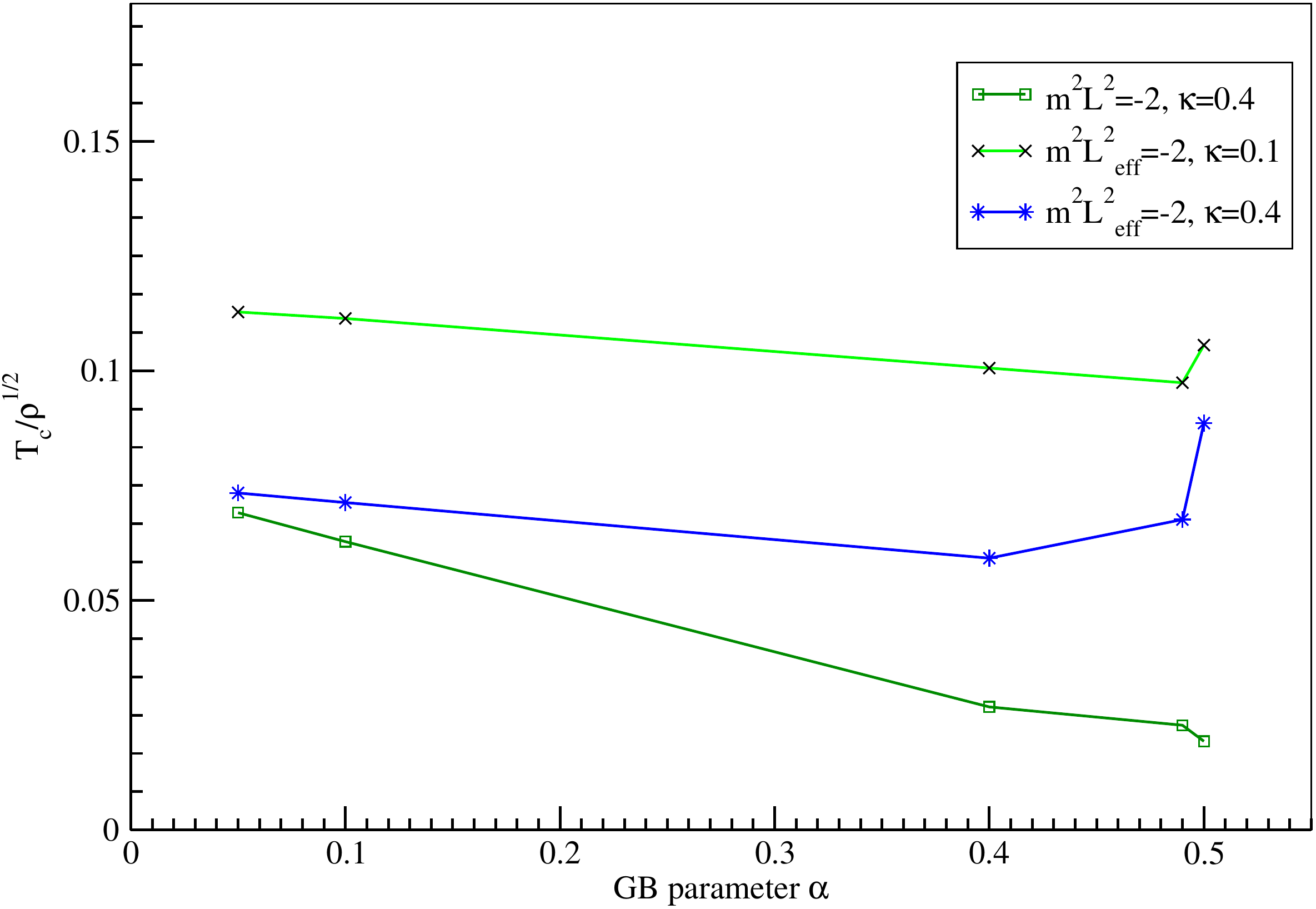} 
\caption{Plot of $\frac{T_c}{\sqrt{\rho}}$ vs. GB parameter $\alpha$ for $m^2L^2=-2$ and $m^2L^2_{eff}=-2$.}
\label{fig2}
\end{figure}

\noindent We have summarized our analytical result of the critical temperature with the GB parameter in Figure \ref{fig2}. We have observed the effect of GB coupling parameter and the backreaction parameter on the critical temperature. For fixed value of backreaction parameter (or GB coupling parameter), the critical temperature decreases with increase in the GB parameter (or backreaction parameter) except at the Chern-Simons limit of the GB coupling value $\alpha=0.5$. We have also displayed that the effects of the backreaction parameter and GB parameters in Figure \ref{fig2}.

%%%%%%%%%%%%%%%%%%%%%%%%%%%%%%%%%%%%%%%%%%%%%%%%%%%%%%%%%%%%%%%%%%%%%%%%%%%%%%%%%%%%%%%%%%%%%%%%%%%%
%%%%%%%%%%%%%%%%%%%%%%%%%%%%%%%%%%%%%%%%%%%%%%%%%%%%%%%%%%%%%%%%%%%%%%%%%%%%%%%%%%%%%%%%%%%%%%%%%%%%

\section{The condensation operator: Critical exponent}
In this section, we shall investigate the effect of the backreaction on the condensation operator near the critical temperature for $4d$ GB gravity. To investigate the critical exponent, we write the field eq.(\ref{new5}) in $z$-coordinate near the critical temperature $(T_{c})$ substituting eq.(\ref{sol1}). This leads to  
\begin{eqnarray}
\phi^{\prime \prime}(z)  = \frac{\langle\mathcal{O}_{+}\rangle^{2}}{r^{2}_{+}} \mathcal{B}(z)\phi (z)
\label{cx1} 
\end{eqnarray}
where $\mathcal{B}(z)= \frac{2 z^{2\Delta_{+} -4}}{r_{+}^{2\Delta_{+} -4}}\frac{F^{2}(z)}{f(z)}~.$ \\
We have considered here $f(z)$ as the GB gravity metric. We may now expand $\phi(z)$ in the small parameter $\frac{\langle\mathcal{O}_{+}\rangle^{2}}{r^{2}_{+}}$ as 
\begin{eqnarray}
\frac{\phi(z)}{r_{+}} = \lambda  (1-z)  + \frac{\langle\mathcal{O}_{+}\rangle^{2}}{r^{2}_{+}} \zeta (z)
\label{cx2} 
\end{eqnarray}
with $\zeta (1)= 0 =\zeta^{\prime}(1).$\\
Using eq.(\ref{cx2}) and comparing the coefficient of $\frac{\langle\mathcal{O}_{+}\rangle^{2}}{r^{2}_{+}}$ of left hand side and right hand side of the eq.(\ref{cx1}), we get the equation for the correction $\zeta(z)$ near to the critical temperature
\begin{eqnarray}
\zeta^{\prime \prime}(z)  = \lambda \frac{2 z^{2\Delta_{+} -4}}{r_{+}^{2\Delta_{+} -4}}\frac{F^{2}(z)}{f(z)}\mathcal{A}_{1} (z)
\label{cx3}
\end{eqnarray}
where $\mathcal{A}_{1} (z) = (1-z) $. \\
We rewrite the above equation as
\begin{eqnarray}
\frac{d}{dz}\left( \zeta^{\prime}(z) \right) &=& \lambda \frac{2 z^{2\Delta_{+} -2}}{r_{+}^{2\Delta_{+} -2}}\frac{ F^{2}(z)}{g_{0}(z)}   \mathcal{A}_{1} (z) ~.
\label{cx4}
\end{eqnarray}
Using the boundary condition of $\zeta(z)$, we integrate (\ref{cx4}) between the limits $z=0$ and $z=1$. This gives
\begin{eqnarray}
\zeta^{\prime}(z)\mid_{z\rightarrow 0} = -\frac{\lambda}{r^{2\Delta_{+} -2}_{+}} \mathcal{A}_{2}
\label{cx5}
\end{eqnarray}
where $\mathcal{A}_{2} = \int^{1}_{0} dz \frac{2 z^{2\Delta_{+}-2} F^{2}(z)}{g_{0}(z)} \mathcal{A}_{1} (z) $.\\
But from eq.(\ref{cx2}), we get the asymptotic behaviour (near $z=0$) of $\phi(z)$. Comparing the both equations of $\phi(z)$ about $z=0$, we obtain 
\begin{eqnarray}
\mu -\frac{\rho}{r_{+}}z &=& \lambda r_{+} (1-z)  + \frac{\langle\mathcal{O}_{+}\rangle^{2}}{r_{+}} \left\{\zeta(0)+z\zeta^{\prime}(0)+.........\right\}
\label{cx7}
\end{eqnarray}
Comparing the coefficient of $z$ on both sides of eq.(\ref{cx7}), we obtain
\begin{eqnarray}
-\frac{\rho}{r_{+}} = -\lambda r_{+} + \frac{\langle\mathcal{O}_{+}\rangle^{2}}{r_{+}}\zeta^{\prime}(0) ~.
\label{cx8}
\end{eqnarray}
Using eq.(\ref{cx5}), we obtain the relation between the charge density $(\rho)$ and the condensation operator $(\langle\mathcal{O}_{+}\rangle)$ 
\begin{eqnarray}
\frac{\rho}{r^2_{+}} = \lambda\left[1 + \frac{\langle\mathcal{O}_{+}\rangle^{2}}{r^{2\Delta_{+}}_{+}} \mathcal{A}_{2} \right] ~.
\label{cx9}
\end{eqnarray}  
Using the expression for the critical temperature and the definition of $\lambda$ and simplifying eq.(\ref{cx9}), we get
\begin{eqnarray}
\langle\mathcal{O}_{+}\rangle^{2} = \frac{(4\pi T_{c})^{2\Delta_{+}}}{\mathcal{A}_{2}[3- \frac{1}{2}\kappa^2_{i} \lambda^2_{\kappa_{i-1}}]^{2\Delta_{+}}}.\left(\frac{T_{c}}{T}\right)^{2} \left[1- \left(\frac{T}{T_{c}}\right)^{2} \right] ~.
\label{cx10}
\end{eqnarray}
Using the fact that $T \approx T_{c}$, we can write
\begin{eqnarray}
\left(\frac{T_{c}}{T}\right)^{2} \left[1- \left(\frac{T}{T_{c}}\right)^{2} \right] \approx 2\left[1- \left(\frac{T}{T_{c}}\right)\right]
\label{cx11}
\end{eqnarray} 
Using this we obtain the relation between the condensation operator and the critical temperature in general dimension
\begin{eqnarray}
\langle\mathcal{O}_{+}\rangle = \beta T^{\Delta_{+}}_{c} \sqrt{1-\frac{T}{T_{c}}}
\label{cx12}
\end{eqnarray}
where $\beta = \sqrt{\frac{2}{\mathcal{A}_{2}}} \left[\frac{4\pi}{3- \frac{1}{2}\kappa^2_{i} \lambda^2_{\kappa_{i-1}}}\right]^{\Delta_{+}}.$\\

\noindent From this eq.(\ref{cx12}), we find that the critical exponent is $1/2$ which is not affected by the GB gravity, BI coupling parameter and backreaction. 
The choice for $m^2 L^2=-2$ or $m^2L^2_{eff}=-2$ yields $\Delta_{+}=2$. Eq.(\ref{cx12}) then becomes 
\begin{eqnarray}
\langle\mathcal{O}_{+}\rangle = \beta T^{2}_{c} \sqrt{1-\frac{T}{T_{c}}} ~.
\label{cx13}
\end{eqnarray}
Now $\mathcal{A}_{2}$ and $\beta$ become 
\begin{eqnarray}
\mathcal{A}_{2} = \int^{1}_{0} dz \frac{2 z^{2} F^{2}(z)}{g_{0}(z)} (1-z) ~~~:~~~ \beta = \sqrt{\frac{2}{\mathcal{A}_{2}}} \left[\frac{4\pi}{3- \frac{1}{2}\kappa^2_{i} (\lambda^2|_{\kappa_{i-1}} )} \right]^{2} ~.
\label{cx14}
\end{eqnarray}
where $F(z)=1-\tilde{\alpha} z^2$ with parameter $\tilde{\alpha}$ at which the  Sturm-Liouville eigenvalue $\lambda^2$ is minimum. We have shown the condensation operator value without backreaction for $m^2L^2=-2$ and $m^2L^2_{eff}=-2$ in Table \ref{tab5}.
\begin{table}[h!]
\caption{The value of $\beta=\frac{\langle\mathcal{O}_{+}\rangle}{T^{2}_{c} \sqrt{1-\frac{T}{T_{c}}}}$ for $\kappa=0$}
\centering
\begin{tabular}{|c| c| c| c| c| c| c| c| }
\hline
$\beta$ & $\alpha=-0.1$ & $\alpha=0$& $\alpha=0.1$ & $\alpha=0.2$ & $\alpha=0.3$ & $\alpha=0.4$ & $\alpha=0.5$  \\
\hline
$m^2L^2=-2$ & 65.10 & 65.62 & 66.08 & 66.87 & 66.27 & 64.87 & 52.88 \\
\hline
$m^2L^2_{eff}=-2$ & 64.86 & 65.62 & 66.41 & 67.17 & 67.78 & 67.72 & 60.98 \\
\hline
\end{tabular}
\label{tab5}
\end{table} 
With backreaction, the values of condensation operator are shown for $m^2L^2=-2$ and $m^2L^2_{eff}=-2$ in Table \ref{tab6} and Table \ref{tab7} respectively. It is observed that the condensation operator value increases first with increasing value of GB coupling value $\alpha$, then the condensation operator value decreases in the near of Chern-Simon limit of $\alpha$. We also observe that the condensation operator value decreases only with higher value of $\alpha$ for fixed higher value of backreaction parameter $\kappa$. 
\begin{table}[h!]
\caption{The value of $\beta=\frac{\langle\mathcal{O}_{+}\rangle}{T^{2}_{c} \sqrt{1-\frac{T}{T_{c}}}}$ for different values of $\kappa$ and $\alpha$ in the case $m^2L^2=-2$}
\centering
\begin{tabular}{|c| c| c| c| c| }
\hline
$\beta$ & $\alpha=0.05$ & $\alpha=0.10$ &  $\alpha=0.40$ &  $\alpha=0.50$   \\
\hline
$\kappa=$ 0.1 & 65.10 & 65.17 & 61.32 & 44.75 \\
\hline
$\kappa=$ 0.2 & 62.66 & 62.26 & 49.58 & 23.16  \\ 
\hline
$\kappa=$ 0.3 & 58.36 & 57.08 & 31.24 & 12.16  \\
\hline
$\kappa=$ 0.4 & 52.01 & 49.40 & 18.02 & 6.55  \\
\hline
\end{tabular}
\label{tab6}
\end{table} \\
\begin{table}[h!]
\caption{The value of $\beta=\frac{\langle\mathcal{O}_{+}\rangle}{T^{2}_{c} \sqrt{1-\frac{T}{T_{c}}}}$ for different values of $\kappa$ and $\alpha$ in the case $m^2L^2_{eff}=-2$}
\centering
\begin{tabular}{|c| c| c| c| c| }
\hline
$\beta$ & $\alpha=0.05$ & $\alpha=0.10$ &  $\alpha=0.40$ &  $\alpha=0.50$   \\
\hline
$\kappa=$ 0.1 & 65.30 & 65.63 & 66.01 & 59.07 \\
\hline
$\kappa=$ 0.2 & 63.07 & 63.18 & 60.68 & 53.66  \\ 
\hline
$\kappa=$ 0.3 & 59.16 & 58.91 & 52.18 & 47.47  \\
\hline
$\kappa=$ 0.4 & 53.44 & 52.71 & 42.39 & 42.78  \\
\hline
\end{tabular}
\label{tab7}
\end{table} \\

\section{Conclusions}
The effect of higher curvature corrections on $3+1$-dimensional holographic superconductors has been investigated analytically in the framework of $4d$ Einstein-Gauss-Bonnet gravity away from the probe limit using Sturm-Liouville eigenvalue method. We have calculated the critical temperature and the condensation operator values for this holographic superconductor model. In this study, we observe that the critical temperature decreases with increasing value $\alpha$ for both cases $m^2L^2=-2$ and $m^2L^2_{eff}=-2$. However, the critical temperature near the upper bound of GB parameter $\alpha$ suddenly increases for $m^2L^2_{eff}=-2$, and this is quite different from the case $m^2L^2=-2$.  Our analytical results match wonderfully with the numerical findings in \cite{jing}. We also observe that higher values of the GB and the backreaction parameters disfavour the condensation since the critical temperature decreases with increasing value of both parameters except at the Chern-Simons limit of the GB parameter. In future we are planning to study the effects of higher curvature correction on the critical value of the magnetic field in $3+1$-dimensional holographic superconductors models.

\section*{Acknowledgments} DG would like to thank DST-INSPIRE for financial support.

%%%%%%%%%%%%%%%%%%%%%%%%%%%%%%%%%%%%%%%%%%%%%%%%%%

\end{document}